\documentclass[a4paper,11pt]{article}
\pdfoutput=1 
\usepackage{jcappub, bm, color} 
\usepackage{amssymb,amsfonts,slashed,amsthm,amsmath,graphicx, soul, empheq, cancel}
\usepackage[caption=false]{subfig}
\begin{document}

\renewcommand{\figurename}{Fig.}
\renewcommand{\tablename}{Table.}
\newcommand{\Slash}[1]{{\ooalign{\hfil#1\hfil\crcr\raise.167ex\hbox{/}}}}
\newcommand{\bra}[1]{ \langle {#1} | }
\newcommand{\ket}[1]{ | {#1} \rangle }
\newcommand{\beq}{\begin{equation}}  \newcommand{\eeq}{\end{equation}}
\newcommand{\bef}{\begin{figure}}  \newcommand{\eef}{\end{figure}}
\newcommand{\bec}{\begin{center}}  \newcommand{\eec}{\end{center}}
\newcommand{\non}{\nonumber}  \newcommand{\eqn}[1]{\begin{equation} {#1}\end{equation}}
\newcommand{\laq}[1]{\label{eq:#1}}  
\newcommand{\dd}[1]{{d \o d{#1}}}
\newcommand{\Eq}[1]{Eq.~(\ref{eq:#1})}
\newcommand{\Eqs}[1]{Eqs.~(\ref{eq:#1})}
\newcommand{\eq}[1]{(\ref{eq:#1})}
\newcommand{\Sec}[1]{Sec.\ref{chap:#1}}
\newcommand{\ab}[1]{\left|{#1}\right|}
\newcommand{\vev}[1]{ \left\langle {#1} \right\rangle }
\newcommand{\bs}[1]{ {\boldsymbol {#1}} }
\newcommand{\lac}[1]{\label{chap:#1}}
\newcommand{\SU}[1]{{\rm SU{#1} } }
\newcommand{\SO}[1]{{\rm SO{#1}} }
\def\({\left(}
\def\){\right)}
\def\dt{{d \o dt}}
\def\diag{\mathop{\rm diag}\nolimits}
\def\Spin{\mathop{\rm Spin}}
\def\O{\mathcal{O}}
\def\U{\mathop{\rm U}}
\def\Sp{\mathop{\rm Sp}}
\def\SL{\mathop{\rm SL}}
\def\tr{\mathop{\rm tr}}
\def\ebq{\end{equation} \begin{equation}}
\newcommand{\OR}{~{\rm or}~}
\newcommand{\AND}{~{\rm and}~}
\newcommand{\EV}{ {\rm \, eV} }
\newcommand{\KEV}{ {\rm \, keV} }
\newcommand{\MEV}{ {\rm \, MeV} }
\newcommand{\GEV}{ {\rm \, GeV} }
\newcommand{\TEV}{ {\rm \, TeV} }
\def\o{\over}
\def\a{\alpha}
\def\b{\beta}
\def\c{\varepsilon}
\def\d{\delta}
\def\e{\epsilon}
\def\f{\phi}
\def\g{\gamma}
\def\h{\theta}
\def\k{\kappa}
\def\l{\lambda}
\def\m{\mu}
\def\n{\nu}
\def\p{\psi}
\def\q{\partial}
\def\r{\rho}
\def\s{\sigma}
\def\t{\tau}
\def\u{\upsilon}
\def\w{\omega}
\def\x{\xi}
\def\y{\eta}
\def\z{\zeta}
\def\D{\Delta}
\def\G{\Gamma}
\def\F{\Phi}
\def\P{\Psi}
\def\S{\Sigma}
\def\me{\mathrm e}
\def\ol{\overline}
\def\tl{\tilde}
\def\*{\dagger}
\def\H{H_{\rm ubble}}

\newcommand{\del}{\partial} 



\title{
Asymmetric Warm Dark Matter: from Cosmological Asymmetry to  Chirality of Life
}

\author{
Wen Yin$^{1,2}$, Shota Nakagawa$^{3,4}$,
Tamaki Murokoshi$^{5}$, \\and Makoto Hattori$^{5}$
}

\affiliation{$^1$ Department of Physics, Tokyo Metropolitan University, Tokyo 192-0397, Japan} 
\affiliation{$^2$ Department of Physics, Tohoku University, Sendai, Miyagi 980-8578, Japan} 

\affiliation{
$^3${ Tsung-Dao Lee Institute, Shanghai Jiao Tong University, \\
No.~1 Lisuo Road, Pudong New Area, Shanghai, 201210, China }
}

\affiliation{
$^4${ School of Physics and Astronomy, Shanghai Jiao Tong University, \\
800 Dongchuan Road, Shanghai 200240, China}
}

\affiliation{$^5$ Astronomical Institute, Tohoku University, Sendai, Miyagi 980-8578, Japan}

\abstract{
We investigate a novel scenario involving asymmetric keV-range dark matter (DM) in the form of right-handed (sterile) neutrinos. Based on the Fermi-Dirac distribution, we demonstrate that asymmetric fermionic DM forms a Fermi degenerate gas, making it potentially colder than symmetric fermionic DM. This setup simultaneously accounts for the Universe's baryon asymmetry through tiny Yukawa interactions with Standard Model leptons and the Higgs field, and the homochirality of amino acids via decay into circularly polarized photons. This scenario can be investigated through soft X-ray searches conducted by current and upcoming space missions.
The helical X-rays is a smoking-gun signal of our scenario. Additionally, we propose a new mechanism to suppress DM thermal production by introducing a light modulus, which may also benefit cosmology involving generic right-handed neutrinos with large mixing.
}

\emailAdd{wen@tmu.ac.jp}

\maketitle
\flushbottom

\section{Introduction
\label{introduction}}

Despite the near symmetry in particle theory, our Universe exhibits significant asymmetries, such as baryon asymmetry (necessary for the origin of matter) and terrestrial bioorganic homochirality. They are essential for the origin of life, yet their origins remain elusive. Another significant enigma across particle theory, astrophysics, and cosmology is the nature of dark matter (DM), which also plays a crucial role for the origin of life, as our Milky Way galaxy would likely be absent without it.

In particle theory and cosmology, baryon asymmetry and DM are key subjects of discussion and study. The hypothesis of asymmetric DM has been proposed~\cite{Barr:1990ca, Kaplan:1991ah, Kitano:2004sv, Kaplan:2009ag} to explain the baryon asymmetry and DM simultaneously. 
The terrestrial bioorganic homochirality has not been directly related to these mysteries. One of the most attractive hypothesis~\cite{Bonner-Rubenstein} suggests that homochirality arises from circularly polarized photons. 

In this paper, we propose a unified explanation for the first time.
We point out that DM, baryon asymmetry, and biological homochirality, three crucial components for life, may all originate from a single particle: asymmetric warm DM (AWDM), which is a right-handed neutrino (RHN)-like Dirac fermion. 
The fermionic DM is produced in the early Universe with significant asymmetry, potentially through the decay of an Affleck-Dine field~\cite{Affleck:1984fy,Murayama:1993em,Dine:1995kz}. Due to the tiny Yukawa interactions with the Standard Model (SM) lepton and Higgs fields, this asymmetry slightly transfers to the lepton asymmetry, which then transforms into baryon asymmetry via sphaleron processes during the electroweak phase transition similar to the leptogenesis scenario~\cite{Fukugita:1986hr}.
The large asymmetry carried by the DM leads to a fermion chemical potential exceeding the temperature, forming a Fermi degenerate gas and thus complying with constraints from small-scale structure formation despite its mass being around or slightly smaller than keV.
In the contemporary Universe, this AWDM decays into a photon and a neutrino through the neutrino mixing resulting from the same Yukawa interaction. Owing to its asymmetry and chirality, the resulting photon is helical and can interact with amino acids in a chiral manner.
We also perform an estimation of the required helical photons for this to naturally happen in a generic context, finding that the parameter region explaining all three origins is consistent (see Fig.\ref{fig:0} for a schematic description of the cosmology). Therefore, the origin of life may rely on a single hypothetical `particle'—AWDM. This scenario can be probed by X-ray observations in the missions, such as Athena, eROSITA, Theseus and XRISM. We consider it is crucial to measure the helicity of the resulting X-rays, which is the smoking-gun signal of our scenario, in the future.
We also show that if we introduce  additional Majorana neutrinos with similar coupling and mass as the AWDM, the neutrino mass can be simultaneously explained through the seesaw mechanism, surprisingly. 
In addition, we point out a new simple mechanism for avoiding the thermalization of the RHN with large mixing.

\begin{figure}[t!]
\begin{center}  
\includegraphics[width=155mm]{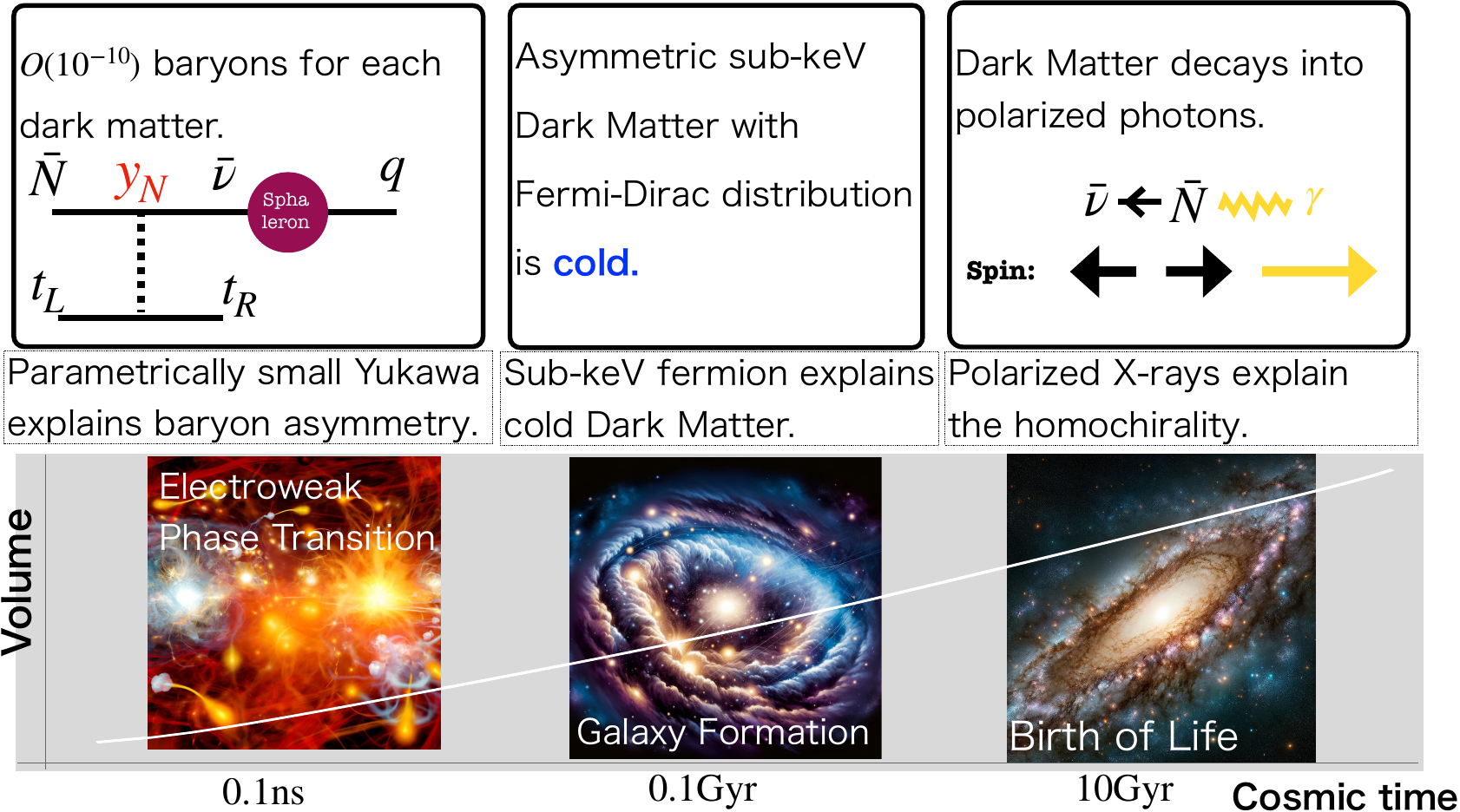}
\end{center}
\caption{A schematic description of the cosmology of AWDM.}
\label{fig:0}
\end{figure}

Some relevant studies are as follows.
In \cite{Cline:2020mdt}, the authors explored a reheating/inflation scenario where the inflaton is identified as the Affleck-Dine field in the Affleck-Dine mechanism. Their study involves the quasi-Dirac RHN to facilitate baryogenesis. In their model, the excessive baryon asymmetry is partially neutralized because the RHN exhibits partial Dirac characteristics.
In contrast, in our scenario, the RHN DM is fully Dirac-type, enabling the generation of helical photons. The resulting baryon asymmetry is significantly less than $\O$(1) due to the small Yukawa coupling of the RHN. These aspects, along with the coldness of asymmetric fermionic DM and its connection to the homochirality of amino acids or/and the active neutrino mass, have not been addressed in previous studies.

It was discussed in \cite{Kasai:2024diy} that large asymmetry of the active neutrinos can be generated due to the Affleck-Dine mechanism and Q-ball decay. Then the RHN (or sterile neutrino) DM (see Refs.\,\cite{Drewes:2016upu,Boyarsky:2018tvu} for review) is produced through the Shi-Fuller mechanism~\cite{Shi:1998km}. 
This differs from our scenario, where RHN DM is directly produced from the decay of the Affleck-Dine field.

The rest of this paper is organized as follows.
In Sec.\ref{sec:homochirality}, we clarify the required abundance of circularly polarized photons after a brief review on homochirality of amino acids, and present our idea of decaying DM explanation.
In Sec.\ref{sec:fermionDM}, we introduce the particle model of AWDM and show the parameter region allowed for the origin of homochirality and baryon.
In Sec.\ref{sec:suppression}, the  overproduction problem is discussed, and in Sec.\ref{sec:neutrinomass}, we give a unified explanation for active neutrino mass. 
Sec.\ref{sec:conclusions} is devoted to the discussion and conclusions. In Appendix \ref{chap:AD}, the Affleck-Dine mechanism is reviewed as an example for inducing large asymmetry of AWDM. In Appendix \ref{chap:opposite}, the suppression mechanism is discussed in a different parameter region.

\section{Homochirality of amino acids and DM explanation
\label{sec:homochirality}}

It is well known that there is an enantiomeric excess of L-form amino acids in proteins, a phenomenon called terrestrial bioorganic homochirality\footnote{The terminology ``homochirality" usually includes D-form sugar domination. Although our conclusion may be applied to it, we focus on the amino acids in this paper.} (see \cite{Bonner_1991,sym11070919} for reviews).
It is believed that the homochirality is important for the origin of life. 
Since organic compounds produced in laboratories show no preference for L- or D-forms, their origin must be explained by some mechanism.
One of the possible hypotheses relies on the geological way in which chiral minerals, e.g. quartz, preferentially absorb only one form of amino acids \cite{Bonner_1976,Furuyama_2006}.
Another possibility is based on symmetry breaking in nature.  
For example, it has been suggested that circularly polarized photons can selectively degrade D-form amino acids (see \cite{sym11070919,Bonner_1991} and references therein).
One of the most attractive recent hypotheses is the following cosmic scenario in which bioorganic molecules are exposed to circularly polarized photons in cosmic space:
\begin{itemize}
\item[(i)] Bioorganic molecules with no enantiomeric excess exist on the surfaces of interstellar dust in star-forming regions \cite{Greenberg1995}.
\item[(ii)] Circularly polarized photons produce an asymmetry between L- and D-forms, creating `asymmetry seeds' including bioorganic precursors with small but non-negligible enantiomeric excess as well as the organic materials, such as the amino acids themselves.
\item[(iii)] Meteorites or asteroids deliver them to Earth, and some chemical or biological mechanism amplifies the transported ``asymmetry seeds" on the Earth.
\end{itemize}
{In step (ii), several sources of such photons were discussed, such as supernove/neutron stars~\cite{Bonner-Rubenstein,Bonner_1991}, dust grains~\cite{Gledhill2000,Bailey_1998,Kwon_2013,Fukushima2023}, and interstellar Lyman $\alpha$~\cite{Hori_2022,Shoji_2023,Fukushima2023}. It was pointed out that the required circularly polarized photon spectrum should be likely a narrow line~\cite{Fukushima2023} because the sign of circular dichroism alternates in sign and sums to zero over the whole spectrum due to the Kuhn-Condon sum rule~\cite{TF9302600293,RevModPhys.9.432}. 
All the previous study require the locally polarized photons while they are not polarized averaged over a wide range.}  
A supportive source of step (iii) is the observation of various kinds of organic compounds in carbonaceous chondrites, especially the Murchison meteorite contains enantiomeric excesses of amino acids \cite{Cronin1997,Engel1997}.\footnote{The amount of delivered organic carbon on the early Earth was estimated from the infall rate of meteorites \cite{Chyba1992}. 
The stability of amino acids at the impact of meteorites on the atmosphere or ground has been also studied experimentally \cite{BASIUK2001231}, but it is still unknown how amino acids are pyrolized in the real Universe.}
The amplification process in step (iii) is ambiguous, but one plausible mechanism is the Soai reaction \cite{Soai2008}, an example of asymmetric autocatalysis, which can amplify a small enantiomeric excess to a significant asymmetry.
We note that there are many uncertainties in the above scenario itself. One uncertainty is the lack of knowledge about when and where the asymmetry generation occurs. Another is the amount of asymmetry that must be induced in cosmic space. These uncertainties should be addressed from various viewpoints including physics, astronomy, biology, and chemistry, and the improvement is expected in the future. 
In this paper, we will proceed our discussion with taking account of the uncertainties.

In \Sec{1}, we estimate the required time-averaged flux of the circulary polarlized photons in  the step (ii), because we could not find a reference for estimating this. 
In \Sec{2}, we study a novel scenario that the circularly polarized photons are generated by asymmetric DM and estimate the required DM lifetime.

\subsection{Asymmetry seeds and flux of circularly polarized photons}
\lac{1}

Let us estimate the required amount of circularly polarized photons for the predominance of L-amino acids based on current experiments.
Indeed, the measurement and analysis of natural circular dichroism in Ref.\,\cite{Yudai_Izumi_2009} show that the difference in absorption cross-section between left- and right-handed circularly polarized photons by L-serine (Ser) and L-alanine (Ala) is estimated to be
\beq 
\D\s(E_\chi) \equiv \s_L(E_\chi)-\s_R(E_\chi) \sim -10^{-21}\rm cm^2
\eeq 
for irradiated photon energy $E_\chi\sim 533$ eV. 
$\D\s$ is normalized such that the cross-section for D-Ser and D-Ala has the same magnitude but the opposite sign. 
Although the cross-section for L-Ala is about $10^{-20} \text{cm}^2$, we conservatively use the result for L-Ser.
When either left- or right-handed circularly polarized light is dominant in the cosmic space, we can consider $|\D\s|$ as the probability of an enantiomeric excess to be generated.

Although the cross-section in the other energy range is still being studied experimentally,\footnote{We thank Junichi Takahashi and Kensei Kobayashi for the information and discussion.}
we use this cross-section as an order-of-magnitude estimate for other photon energy ranges because the cross-section between amino acids and photons should not differ significantly from that of typical molecules. 
In \cite{Yudai_Izumi_2009}, D-Ser and D-Ala are erased by left-handed circularly polarized photons. 
However, we cannot conclude which polarization of the photon is required for biasing enantiomers at the moment. 
This is because the answer strongly depends on the type of amino acids and experimental conditions as well as the energy of photons~\cite{TF9302600293,RevModPhys.9.432,Fukushima2023}, and further experiments and theoretical explorations are needed.

Therefore, we proceed with our discussion without specifying the polarization of the produced circularly polarized light.

Following various studies in this area~\cite{Takahashi_1999_abio,Yudai_Izumi_2009,SOAI2022133017,Hori_2022,Shoji_2023,Fukushima2023,ASato_2023},  we assume either that the relevant amino acids are very long-lived on cosmological time scales or that they are not so long-lived, but through continuous creation and destruction, the total number remains approximately constant.
In the latter case, we consider that homochirality is preserved; for example, once an asymmetry seed is generated, it is maintained during subsequent creation and destruction processes (In other words, we consider the case that the racemization is neglected).
Then, the difference in the interaction rate between each enantiomer of an amino acid and circularly polarized photons is estimated as
\beq 
\Gamma_\chi =\Phi_\chi |\D\s| \sim (3{\rm Gyr})^{-1} \frac{\Phi_\chi}{10^{-9}\times c n_{\rm CMB}}.
\eeq 
Here, $\Phi_\chi$ is the flux of the circularly polarized photon, and we denote the current number density of CMB photons as $n_{\rm CMB} = (2\zeta(3)/\pi^2)T_0^3$ with $T_0 \approx 0.23\,$meV as the present temperature of CMB photons and $\zeta(3)\simeq1.2020$ as the zeta function. $c$ is the speed of light, which will be taken to be unity in our natural unit. 
This means that if the number density (or flux) of circularly polarized photons is $10^{-9}$ times larger than that of CMB photons, amounting to $\O(10^4)$ photons $[\rm cm^{-2} s^{-1}]$, then the enantiomeric excess can be produced within $3\,$Gyr. 
This also roughly represents the required time-averaged photon flux for a generic cosmic scenario.
For a time-averaged photon flux smaller than this, it is improbable for the reaction to occur within the timescale for the birth of life, even if there is a rare event with a significant circularly polarized photon production.\footnote{Satisfying this condition the photon flux from the DM decay can be also more significant in a short period due to the probability fluctuation.}

\subsection{Asymmetry seeds from decaying DM} 
\lac{2}

Now, we consider the possibility that DM decays into circularly polarized photons to induce homochirality.
The particle model for the circularly polarized photons will be discussed in the next section.
The differential photon flux from the DM decay in our Milky Way galaxy is given by
\beq
\frac{\partial \Phi_{\chi}}{\partial E_\chi}=\int d\Omega dl  \frac{1}{4\pi l^2}   \left(\frac{\Gamma_{\rm DM}  \rho_{\rm DM}(l, \Omega)}{m_{\rm DM}}\right) \, l^2\, \frac{\partial^2 N_{\chi}}{\partial E_\chi \partial \Omega }[l, \Omega],
\eeq
where $\frac{\partial^2 N_{\chi}}{\partial E_\chi \partial \Omega }[ l, \Omega]$  denotes the energy spectrum from a single DM decay at the line-of-sight distance $l$ and in the direction  $\Omega$, 
and $\rho_{\rm DM}, m_{\rm DM}$, and $\Gamma_{\rm DM}$ represent the local energy density, mass, and decay rate into photons of DM, respectively.
The integration is taken over the observable Universe.
Assuming that the energy spectrum of the produced photons is homogeneous and isotropic, we can factorize as
\beq 
\frac{\partial \Phi_{\chi}}{\partial E_\chi}= \frac{D}{4\pi} \times  \frac{\G_{\rm DM}}{m_{\rm DM}} \frac{\partial N_{\chi}}{\partial E_\chi}
\eeq 
with 
\beq 
D \equiv \int dl d\Omega \rho_{\rm DM}(l,\Omega).
\eeq 
We approximate $D=(8.1-8.7)\times 10^{-5} \GEV^{3}\simeq \(1.8-1.9\) \times 10^9 M_{\odot} {\rm kpc}^{-2}$, 
by using the DM density profile in our Galaxy (we used the NFW, Einasto profiles in Ref.\,\cite{Cirelli:2010xx}, and the one from Gaia DR3 data in Ref.\,~\cite{Jiao:2023aci} for estimation).\footnote{{We estimated the D-factor around the solar system by assuming that the asymmetry seeds of the amino acids are generated at a similar distance from the galaxy center to that of the solar system. However in reality, depending on the position of the reaction to happen, the estimation may be different. This uncertainty is also included in $\xi_{AA}$.}}
{By requiring the photon flux to be $\xi_{AA}\times 10^{-9}n_{\rm CMB}$, }
this establishes the favored value for the DM decay rate given the mass,
\beq 
N_\chi\G_{\rm DM} \sim \xi_{AA} \times {10^{-25}}{\rm s}^{-1} \frac{m_{\rm DM}}{1\KEV }.
\label{eq:DMdecay}
\eeq 
{Here, $\xi_{AA}$ represents the uncertainty factor from our previous estimation of the photon flux necessary to explain the homochirality of amino acids.}
Interestingly, this decay rate is compatible with established bounds for DM decaying into photons and for the coldness of AWDM.
In particular, when the resulting photon has an energy smaller than $0.5\KEV$, the bound becomes significantly weaker. 
Currently, the most stringent bound for these photon emissions is the heating constraint from the Leo T dwarf galaxy~\cite{Wadekar:2021qae},  which imposes an upper limit of $N_\chi \G_{\rm DM} \lesssim 10^{-25} \rm s^{-1}$.
Future X-ray observations, such as those planned for Athena and Theseus, will probe DM decay within the same mass range e.g. \cite{Bulbul:2014sua}. 
Ongoing missions such as eROSITA and XRISM may also have the opportunity.

An important question is how to produce the circularly polarized photon.
Assuming the total non-relativistic DM in the Milky Way is not polarized,\footnote{For cohererent oscillating vector boson DM, this assumption  may not be satisfied~\cite{Caputo:2021eaa}. We can get a polarized photon from the polarized vector boson decaying into a photon and a scalar field. The polarization and thus the resulting homochirality depends on the place we live in.} charge-conjugation-parity (CP) transformation says that the DM is asymmetric; 
otherwise, the averaged number of the resulting polarized photons is produced at the same rate of that for the CP conjugate one, i.e., the photons with opposite polarization. 
Similarly, we must also violate parity symmetry.
This is achievable if the decay products include a chiral fermion, which inherently violates parity. 
One simple possibility is to consider the fermionic asymmetric DM decaying into a chiral fermion and a photon. 
In the next section, we will examine the asymmetric fermionic DM, and especially, will focus on a RHN DM.

\section{Asymmetric Warm Dark Matter
\label{sec:fermionDM}}

\subsection{Degenerate fermion DM}

Let us give a generic discussion on an asymmetric fermionic DM, denoted by $N$. 
Here we assume that the dark sector is very weakly coupled to the visible sector, and the intrinsic temperature, denoted as $T_{\rm dark}$, is considerably lower than that of the visible sector, $T$.
The DM particle obeys the Fermi-Dirac distribution,
\beq f_N[E,\mu]= \frac{1}{\exp\left(\frac{E-\m}{T_{\rm dark}}\right) + 1},
\eeq
where $\m$ denotes the chemical potential of the DM. As long as the DM is produced asymmetrically and the comoving asymmetry is conserved, this remains the final equilibrium distribution given some interactions for the thermalization.

Note that the chemical potential with the negative sign is assigned to the antiparticle of the DM, $\bar N$, i.e. the distribution is
$f_{\bar{N}}[E,\mu]=f_N[E,-\mu]$.
By neglecting the DM mass, $M_N \ll |\mu|$, we estimate the number density to be
\beq 
n_{N}=-g_{N}\frac{T_{\rm dark}^3}{\pi^2} {\rm Li}_3\left(-e^{\frac{\m}{T_{\rm dark}}}\right),
\eeq 
where ${\rm Li}_k$ is the polylogarithm of order $k$, and $g_{N}$ is the internal degrees of freedom of the DM (minimally, this is $2$ for the fermionic DM).
In the extreme limits of $|\mu|/T_{\rm dark}$, we obtain its approximated forms as follows:
\beq
n_N \simeq \left\{~
\begin{aligned}
    & g_N\(\frac{3\zeta(3)}{4\pi^2} T_{\rm dark}^3+\frac{1 }{12} T_{\rm dark}^2 \m\) ~~(|\m| \ll T_{\rm dark}) \\
    & g_N \frac{\m^3}{6\pi^2} \hspace{4.1cm} (|\m| \gg T_{\rm dark}).
\end{aligned}
\right.
\label{eq:number}
\eeq
In the latter limit, the number density for $\mu<0$ is evaluated to be almost vanishing. 
For $|\m| \gg T_{\rm dark}$, which is our focus, we have the coldest state for the DM particle, called degenerate Fermi gas or Fermi sea. 
Since the chemical potential scales as $\mu \propto a^{-1}$, with $a$ being the scale factor, and $|\mu|$ identified with the Fermi momentum, most fermions become non-relativistic when $|\mu|$ falls below the DM mass due to the expansion of the Universe. 
Although our discussion does not depend on the specific mechanism of asymmetry production, one possibility is coupling $N$ to an `Affleck-Dine field' in a manner similar to the Affleck-Dine mechanism~\cite{Affleck:1984fy,Murayama:1993em,Dine:1995kz} (see Appendix~\ref{chap:AD}).
{For the correct sign of baryon asymmetry as will be discussed and the coldness of DM, we consider
\beq 
\m <0 \AND |\m|\gg T_{\rm dark}.
\eeq 
Namely, $n_{\bar{N}}\simeq g_N |\m|^3/(6\pi^2) \gg n_N\approx 0.$
Notice that the DM  with $|\mu|\ll T_{\rm dark}$, i.e.  the usual warm DM, cannot be as cold as our DM because then $T_{\rm dark}$ is fixed to explain the abundance of the DM. We call the fermionic DM with its abundance explained by fixing $ |\mu|(\gg T_{\rm dark})$ due to the asymmetry, {\it asymmetric warm dark matter (AWDM)}.}

Using the number density in \Eq{number}, the total DM abundance or density parameter can be estimated as follows:
\begin{eqnarray} 
\Omega_{N}h^2 &\equiv& \frac{\rho_{N0}+\rho_{\bar{N}0}}{\rho_ch^{-2}} = M_N\frac{n_N+n_{\bar N}}{s(T)} \frac{s_0}{\rho_ch^{-2}} \label{eq:Nabund} \\
&\approx& 0.05 g_N \frac{M_N}{1\KEV} \(\frac{\ab{\m}}{0.8T}\)^3 \frac{106.75}{g_{\star s}(T)},
\end{eqnarray}
where $\rho_{N(\bar{N})0}$ is the current energy density of $N$ (the anti-particle, $\bar{N}$), $\rho_c
$ is the current critical density with the reduced Hubble constant $h$, and $s(T)=(2\pi^2/45)g_{\star s}(T)T^3$ denotes the entropy density with $g_{\star,s}(T)$ being the degrees of freedom for entropy density. 
The current value of $s$ is denoted by $s_0$. 
{$T$ is the cosmic temperature. We consider at $T$, the fermion comving asymmetry do not change more (i.e. the asymmetry production stops). }
We neglect the contribution from the antiparticles for the dominant DM in this estimation, assuming $T_{\rm dark}\ll |\m|$.
For $|\mu|\sim0.8T$,  the fermion with $M_N\sim1\KEV$ can explain the total abundance of the DM.

Before proceeding to the next discussion, let us mention a lower bound on the DM mass.
As estimated above, the observed abundance of DM can be saturated by the degenerate Fermi gas. 
Fermions obey the Pauli exclusion principle, and the phase-space density is constrained by observations of DM-rich objects, such as dwarf spheroidal galaxies, leading to a limit on the mass.
We quote the so-called Tremaine-Gunn bound ~\cite{Tremaine:1979we,Boyarsky:2008ju} to conservatively constrain the DM mass $M_N$ as \footnote{See also Refs.\,\cite{Moroi:2020has,Carena:2021bqm} for the recast bound for the free-streaming of the DM from the Lyman-$\alpha$ forest, which is comparable to the phase-space bound. As we will see, our scenario can be a late-forming DM, and we cannot simply adopt this bound.
} 

\beq
M_N\gtrsim 0.4\KEV \(\frac{4}{g_{N}}\)^{1/4}.
\eeq 
Here we adopt the bound from the degenerate Fermi gas for half of the degrees of freedom (e.g., for a single Dirac fermion we have $g_N=4$, but we use the bound for 2 degrees of freedom). This is because only the anti-particles contribute to the DM.

We also note that a recent study shows a somewhat weaker phase-space bound~\cite{Alvey:2020xsk}. Their bound for our scenario reads 
$
M_N\gtrsim 0.13\KEV \(\frac{4}{g_N}\)^{1/4}. 
$ 
If this is the lower bound for the DM mass, then the resulting photon energy from the 2-body decay will be $\gtrsim 65\EV$.\footnote{Therefore, if we adopt this bound, the allowed parameter region will be larger than what we will explain. 
We thank Kohta Murase for letting us know the reference.}

\subsection{Model: right-handed neutrino DM}

Let us explore a specific model of RHN (also known as a sterile neutrino), which is a fermion that mixes with the SM neutrinos (active neutrinos), for the realization of the AWDM.
This is similar to the seesaw model~\cite{Minkowski:1977sc,Yanagida:1979as,Ramond:1979py,Gell-Mann:1979vob,Mohapatra:1979ia}, or, specifically, a Dirac seesaw \cite{Roncadelli:1983ty,Roy:1983be,Dick:1999je,Murayama:2002je,Gu:2006dc,Thomas:2005rs,Abel:2006hr,Ma:2014qra}.
The Lagrangian is given by
\beq 
{\cal L}_N = -(M_N \bar{N} N+ y_N \bar{N} \hat P_L L h + {\rm h.c.}).
\eeq 
Here $N$ denotes the Dirac spinor for the DM, $L$ the SM left-handed lepton, $h$ the SM Higgs doublet, $y_N$ the Yukawa coupling, and $\hat{P}_L\equiv (1-\gamma_5)/2$ is the left projection operator. 
{We have suppressed the generation and $SU(2)_L$ indices, and assumed that $N$ has only one generation, for simplicity. 
Importantly, only the right-handed $N$ couples to the left handed $L$. 
We also have left-handed $N$ since left and right-handed $N$ form the Dirac mass $M_N$. In other words, this Lagrangian conserves the lepton number.}

With the electroweak symmetry breaking, the sterile neutrino has a mixing coupling, $\theta\simeq y_Nv/M_N$, with the active neutrino, where $v\simeq173\GEV$ is the Higgs expectation value. 
The RHN couples to the SM sector and it decays decays into an active neutrino and a photon via a quantum loop effect,
\beq
\bar N\to \bar \nu_L +\gamma. 
\eeq 
The decay rate is given by \cite{Pal:1981rm,Barger:1995ty}
\beq 
\Gamma_{N\rightarrow \nu_L\gamma} = \frac{9\a G_F^2}{256 \pi^4} M_N^5 \theta^2\approx 1.4\times 10^{-26} {\rm s}^{-1} \(\frac{M_N}{1\KEV}\)^5 \(\frac{\theta}{0.5\times 10^{-2}}\)^2,
\eeq 
where $\a\approx 1/137$ denotes the fine structure constant and $G_F=1.166\times10^{-5}\GEV^{-2}$ is the weak coupling constant.
One can easily see that the produced photon should be left-handed polarized (namely, the photon motion is opposite to the spin) in the same way with $N$, while it is polarized in the opposite sign when we have the decay of the anti-DM, $\bar{N}$. 
Since we assume this is an asymmetric DM, say the abundance of $\bar N$ dominates over that of $N$, then it predominantly produces the photon with a single polarization, resulting in a polarized photon in total (see the right panel of Fig.~\ref{fig:0}). 
From \Eq{DMdecay}, the mixing coupling $\theta\sim 0.5\times10^{-2}$ is required for producing the enantiomeric excess for $M_N=1\KEV$.
Since we consider $\mu<0$, $\bar N$ is dominant and the resulting photon is right circularly polarized.\footnote{If we consider the positive chemical potential for the RHN $\mu>0$,  $N$ dominates the DM abundance.
Then the produced photons are left circularly polarized. In this case we cannot explain the baryon asymmetry, as the sign is opposite. Instead of explaining the baryon asymmetry, we may consider to produce positive lepton asymmetry after the electroweak phase transition. This scenario may alleviate the tensions in Hubble parameter measurements and explain the helium-4 anomaly since it is the desired sign~\cite{Matsumoto:2022tlr}. 
In this scenario,  the thermal production may be suppressed due to the small reheating temperature.
}

\subsection{Common origin of baryon and life
\label{sec:baryon}}

Now, let us present the exciting possibility that the AWDM discussed previously is also responsible for the baryon asymmetry
$N$ has $\O(1)$ asymmetry, and couples to the left-handed Lepton and Higgs bosons very weakly. 
Through this coupling, the asymmetry of $N$ slightly leaks into the visible sector lepton asymmetry via reactions such as, e.g.,
\beq 
\bar N t \to  \bar \nu_L t
\eeq 
where $t$ represents the top quark.  
This lepton asymmetry is transferred into the baryon asymmetry due to the sphaleron process. 
The total reaction rate is given by~\cite{Besak:2012qm, Hernandez:2016kel, Ghiglieri:2017gjz,Eijima:2019hey}
\beq
\G_N^{\rm th} \simeq 0.01 |y_N|^2  T.
\eeq 
This scales with $a^{-1}$ and decreases slower than the Hubble parameter $H= \sqrt{\frac{g_\star \pi^2}{90}}\frac{T^2}{M_{\rm pl}}\propto a^{-2}$ in the radiation dominated Universe, 
where $g_\star(T)$ is the relativistic degree of freedom for energy density, and $M_{\rm pl}
=2.4\times10^{18}\GEV$ the reduced Planck mass.
Since a unit interaction transfers a lepton asymmetry to the visible sector, the produced lepton asymmetry is estimated as
\begin{align}
\frac{n_{L}}{s}(T=T_{\rm sph})&\simeq \e \frac{n_N+n_{\bar N}}{s} \left.\frac{\G_N^{\rm th}}{H}\right|_{T= T_{\rm sph}}\laq{asym}\\
&\simeq 3\times 10^{-10} \e \frac{1\KEV }{M_N} \(\frac{y_N}{10^{-10}}\)^2.\laq{DMasym}
\end{align}
Here, we use the sphaleron decoupling temperature, $T_{\rm sph}\sim 100\GEV$, in the r.h.s to approximate the integration of the Boltzmann equation until the decoupling. 
In the last approximation, we used $\Omega_{N} h^2\simeq 0.12$ for explaining the observed DM abundance with \Eq{Nabund}. 
For $\e\equiv (n_N-n_{\bar{N}})/(n_N + n_{\bar{N}})\sim -1$, we can explain the baryon asymmetry via the sphaleron process\,\cite{Khlebnikov:1988sr,Harvey:1990qw}. This sign is why we consider $\mu<0.$
{By requiring $\frac{n_L}{s}(T=T_{\rm sph})= \x_{B} \times \(-3\times 10^{-10}\)$,}
the resulting mixing parameter between the sterile and the active neutrino is given by
\beq
\theta \sim \frac{y_N v}{M_N}\sim 0.02\frac{y_N}{10^{-10}}\frac{1\KEV}{M_N}\sim 0.02\sqrt{\x_B} \sqrt{\frac{M_N}{1\KEV}}.
\eeq
 We have introduced $\xi_B$ to account for the uncertainty in the estimation of the baryon asymmetry.

\begin{figure}[t!]
\begin{center}  
\includegraphics[width=135mm]{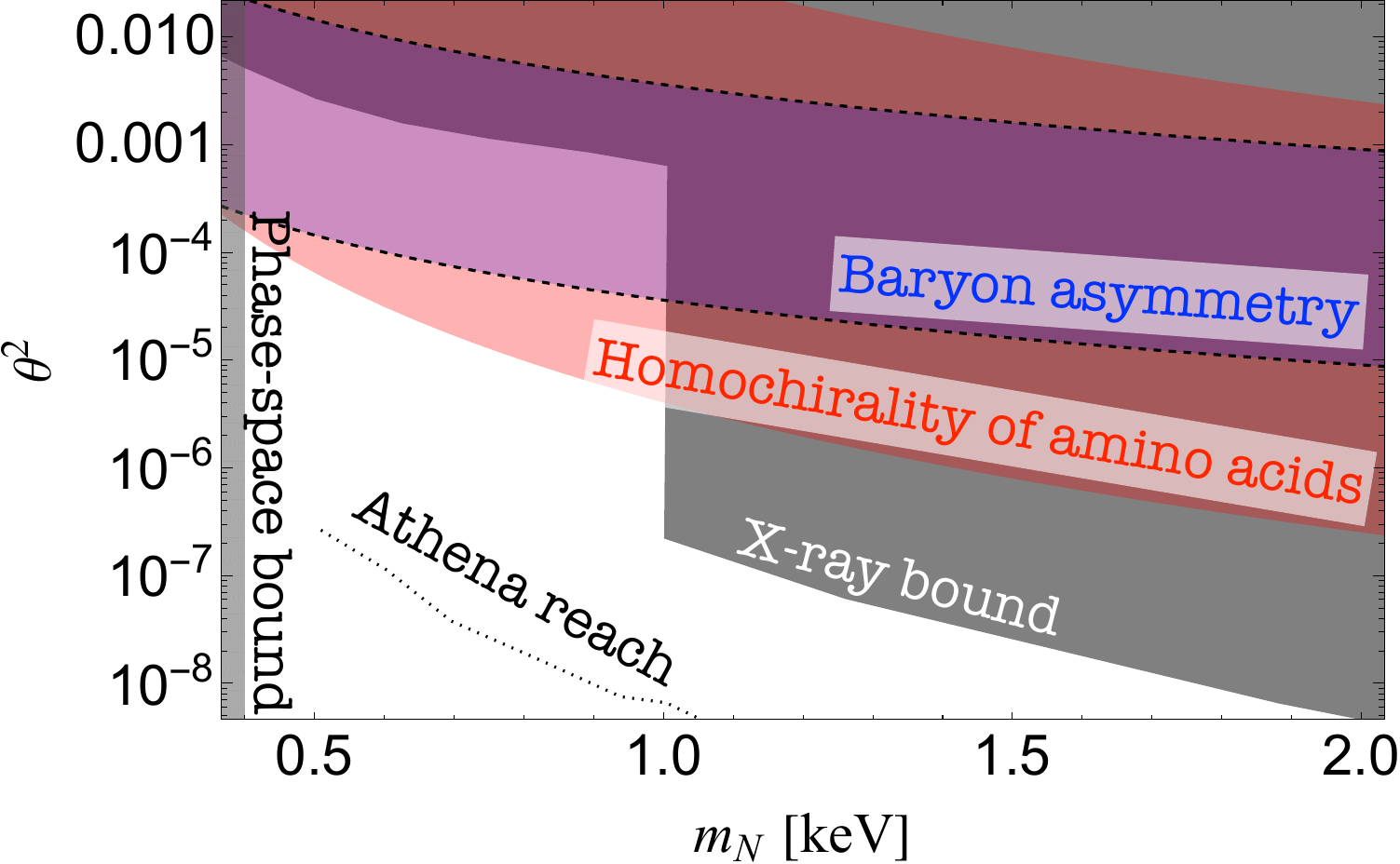}
\end{center}
\caption{The parameter region of the AWDM. The correct baryon asymmetry can be obtained in the blue band. The region favored by the homochirality of the amino acids is represented by the red band. Here we use $\x_{B}=0.1-10$ while $\x_{AA}=0.01-100$. The gray (light gray) region is excluded by the observation relevant to the X-ray emission adopted from \cite{Wadekar:2021qae} (phase-space bound). The dotted line denotes the future reach by Athena~\cite{Neronov:2015kca}.}

\label{fig:1}
\end{figure}

In Fig \ref{fig:1}, we show the parameter region for this AWDM scenario. If the mass is smaller than $1\,$ keV and larger than a few hundred eVs, there exists an interesting parameter region where both the homochirality of amino acids and baryon asymmetry can be explained simultaneously. This region can be probed by searching for X-rays, such as in Theseus, Athena \cite{Thorpe-Morgan:2020rwc}, eROSITA \cite{Dekker:2021bos}, and XRISM\footnote{The XRISM mission faces an issue with the aperture door covering its detector, which has not opened as planned. This door, meant to protect the detector before launch, now blocks lower-energy X-rays, reducing sensitivity to 1.7 keV instead of the intended 0.3 keV. Resolving this issue is crucial for searching for photons from AWDM. 
} \cite{XRISMScienceTeam:2020rvx}. For concreteness, we show the sensitivity reach of Athena. See Refs.\,\cite{Thorpe-Morgan:2020rwc,Dekker:2021bos,Dessert:2023vyl} for  the reaches of Theseus, eROSITA, and XRISM. 
The DM in this parameter region can affect the shape of the de-excitation spectrum for $^{163}$Ho by capturing an electron in ECHo~\cite{Filianin:2014gaa} (see also the HOLMES~\cite{Alpert:2014lfa} and NuMECS~\cite{Engle:2013qka} experiments) and the shape of the tritium $\beta$ decay spectrum in KATRIN~\cite{Angrik:2005ep, Mertens:2014nha, Mertens:2014osa} (see also the Troitsk \cite{Kraus:2004zw}, Project 8~\cite{Asner:2014cwa}, and Ptolemy~\cite{Betts:2013uya} experiments). These experiments will also probe our scenario.

\section{Suppressing the notorious thermal production by a light modulus
\label{sec:suppression}}

So far, we have discussed the non-thermal production of sterile neutrino DM and clarified the condition for explaining baryogenesis and homochirality, which requires a relatively large mixing with the active neutrinos, $\theta\sim\mathcal{O}(10^{-3}-10^{-2})$.
However, in our scenario, we face a potential problem due to such a large mixing interaction. 
When there is a mixing with active neutrinos, thermal production via mixing, based on the so-called Dodelson-Widrow (DW) mechanism~\cite{Dodelson:1993je}, may work, and the produced abundance for $\theta\sim\mathcal{O}(10^{-3}-10^{-2})$ can be excessively high. 
{One possible way is to introduce self-interaction for active neutrinos~\cite{DeGouvea:2019wpf}.\footnote{We studied the self-interaction in the dark sector with the DM rather than the active neutrino sector to induce effective potential of $N$ to suppress the production. However, the production/thermalization via the same interaction is too large to suppress the thermal production sufficiently.}}
In this section, we propose an alternative, novel mechanism to evade this overproduction problem, which applies not only to our scenario but also to other scenarios, such as the sterile neutrino explanation for the short baseline anomaly~\cite{LSND:1996ubh,LSND:1997vun,MiniBooNE:2007uho,MiniBooNE:2010idf,Denton:2021czb,Arguelles:2021meu}.

Let us introduce a real, light modulus field $\phi$, which is a subdominant component throughout the cosmological history.
As we will see shortly, despite its small abundance, it induces an extra time-dependent mass of $N$ and thus the time-dependent mixing. This will naturally suppress the thermal production.
For a simple setup, the interaction of $\f$ with the AWDM is described by the Lagrangian,\footnote{More generically, we can couple the real scalar field to $N$ with arbitrary form without assuming a CP symmetry: 
$\bar N \f(A+i B\g_5) N$. 
It does not change our conclusions much.} 
\beq
{\cal L}_N \supset - y_\phi i \f \bar N  \g_5 N,
\eeq 
where $y_\phi$ denotes the coupling constant.
It is light; therefore, in the early Universe, it is natural that $\f$ is not in the vacuum. Indeed, during inflation the stochastic behavior of $\f$ leads it to the flat distribution~\cite{Starobinsky:1986fx,Starobinsky:1994bd}. After inflation, $\f$ field changes its value due to the dynamics. 
Thus the effective mass of $N$ is time-dependent and is given by 
\beq 
M_{N,t}= \sqrt{M_N^2+(y_\phi \f[t])^2}.
\eeq

After the production of the degenerate Fermi gas, 
the effective mass of $\f$ is obtained because the value of $\f$ affects the free energy of the gas. 
Assuming the $|y_\phi \phi|$ term dominates over $M_N$ but is smaller than $|\mu|$, we get the effective potential
 \beq
V_{\rm eff}\simeq g_N (y_\phi \f)^2 \frac{\m^2}{8\pi^2}.
 \eeq 
In this regime, the effective mass for $\phi$ is 
 \beq 
 m_{\rm eff}= \sqrt{g_N} \frac{y_\phi \ab\m}{2\pi}.
 \eeq

\paragraph{Cosmology of $\f$}

In the following, we assume that the effective mass of $\phi$ dominates over the bare potential of $\f$. 
Note that the effective mass decreases with $a^{-1}$ due to the expansion of the Universe, as $|\mu|$ does.
On the other hand, the Hubble parameter $H$ decreases with $a^{-2}$ in the radiation dominated Universe. Thus, $m_{\rm eff}\gtrsim H$ will be satisfied if the radiation-dominated Universe lasts long enough. 
Suppose $m_{\rm eff}\ll H$ when the asymmetry of $N$ is produced. At this moment, $\f$ is frozen to $\f=\f_{\rm ini}$ due to the Hubble friction.
When $H \sim m_{\rm eff}$ after the redshifts of $H$ and $m_{\rm eff}$, oscillations begin.
We define the temperature as $T\sim T_{\rm osc}$. Then the coupling constant $y_\phi$ is estimated as
\begin{eqnarray} 
y_\phi &\sim& \sqrt{\frac{2}{45}} \pi^2 \sqrt{\frac{g_{\star,{\rm osc}}}{g_N}} \frac{T_{\rm osc}^2}{\ab \m M_{\rm pl}}\\
&\approx& 3.9\times 10^{-16} g_N^{-1/6} g_{\star s,\rm osc}^{-1/3} g^{1/2}_{\star,\rm osc} \(\frac{M_N}{1\KEV}\)^{1/3} \frac{T_{\rm osc}}{100\GEV}
\end{eqnarray}
by replacing $\m$ with $M_N$ from the relation for the observed DM abundance $\Omega_{N}h^2=0.12$. Here we define $g_{\star,{\rm osc}}\equiv g_{\star}(T_{\rm osc})$ and $g_{\star s,{\rm osc}}\equiv g_{\star s}(T_{\rm osc})$.
The coupling is constrained from the DM lifetime of $\bar N\to \bar \nu \f$ with a rate $\sim (y_\phi \theta)^2 M_N/(4\pi)$.
The analysis by using cosmological observations \cite{Enqvist:2019tsa} presented the bound on the lifetime, $\tau_{\rm DM}\geq175{\rm Gyr}$, which gives the upper bound,
\beq
\laq{yphi}
y_\phi \lesssim 2.4\times 10^{-17}\left(\frac{0.05}{\theta}\right)\sqrt{\frac{1\KEV}{M_N}}.
\eeq 
This constraint is easily satisfied if $T_{\rm osc}\lesssim 100\GEV.$\footnote{We also comment that in close proximity to the bound, the resulting active neutrino flux can be higher than the background solar neutrino flux~\cite{Vitagliano:2017odj}, and it is monochromatic.  One may search for this scenario from an indirect detection of the active neutrino. }
Such a small coupling never contributes to the (non-forward) scattering or thermal potential of the DM particles.
Thus these effects will be neglected in the following analysis.

After the onset of oscillation, the adiabatic invariant for $\f$ in comoving volume conserves, and $m_{\rm eff} \bar{\f}^2 \propto a^{-3},$ with $\bar\f$ being the oscillating amplitude of $\f$. 
This implies that the energy density of $\phi$ oscillation scales as radiation,
$
\rho_\f\propto a^{-4}.
$
Then our assumption that the energy density of $\phi$ is always a subdominant component of the Universe can be rephrased by using the initial condition as\footnote{One can also consider a late forming DM scenario, where $\f$ becomes DM.
This is the case that $\f$ has a dominant thermal mass in the early Universe and behaves as radiation, but the bare mass term becomes non-negligible before the cosmic temperature $T>\mathcal{O}(1)\KEV$.
See similar cosmology in Refs.\,\cite{Daido:2017wwb,Daido:2017tbr,Batell:2021ofv, Li:2021fao,Fujita:2023axo}.}
\beq
\frac{m_{\rm eff}^2}{2} \f_{\rm ini}^2 \lesssim g_{\star} \frac{\pi^2}{30} T^4_{\rm osc} ~\leftrightarrow~ \phi_{\rm ini}\lesssim\sqrt{6}M_{\rm pl}.
\eeq 
This is naturally satisfied, if $\f_{\rm ini}$ is not too larger than $M_{\rm pl}.$

Given $\f_{\rm ini}\sim M_{\rm pl}$, we obtain a relation between $y_\phi \bar \f$ and $T$:
\beq
y_\phi \bar\f \simeq 3\frac{ g_{\star,\rm osc}^{1/2}g^{1/3}_{\star s} }{g^{1/6}_N  g^{2/3}_{\star s,\rm osc}} \(\frac{M_N}{1\KEV}\)^{1/3} \frac{\f_{\rm ini}}{M_{\rm pl}} T.
\eeq 
Here, we again use the condition to explain the observed DM abundance to eliminate $\m$. 
This is an interesting consequence, because if $\f_{\rm ini}$ is close to the Planck-scale, which is a natural assumption, we have $y_\phi \bar\f$ of the same order  with, or slightly smaller than the cosmic temperature $T$.
Since we consider $\f$ as a light modulus, we expect the regime of thermal potential domination to last long, and then the large (oscillating) effective mass of $N$ is obtained in the early Universe.

\paragraph{Suppressing the thermal production of $N$}
Here, we discuss the impact of the oscillating effective mass of $N$ on the evolution of the active and sterile neutrino system. 
In the following, we proceed with our analysis in the flavor basis of the neutrino vector  $(\nu,N)$.
Assuming that the neutrinos are relativistic, the Hamiltonian for the active neutrino and a single $N$ with momentum $p$ is given by:
\beq 
\mathcal{H} \simeq p + \frac{M^\dagger\cdot M}{2p}+ V_T.
\eeq 
We denote the thermal potential for the active neutrinos as $V_T=\diag{[V_\nu, 0]}$ with $V_\nu \sim -80 G_F^2 p T^4$ \cite{Venumadhav:2015pla,Gelmini:2019wfp},
and the mass matrix is defined as
\beq 
M =\(\begin{array}{cc}
    0 & y_N v \\
    y_N v &  M_{N,t}
\end{array}\).
\eeq 
Note that we ignore the contributions of the asymmetry of active neutrino. 
Since we assume the tiny coupling $y_N$ with the active neutrinos, the production and destruction of sterile neutrinos are mediated by the coherence-breaking interaction from the active neutrino flavor $\nu$, whose interaction rate (or mean free path) is given by $\G=\diag[\G_\nu,0]$ with $\G_\nu \sim G_F^2 p T^4$ the total thermal width.

Now let us consider the broken phase of electroweak symmetry, with the onset of $\phi$ oscillation occurring before the QCD phase transition.
As will be discussed, if the onset of oscillation is delayed, the suppression of thermal production will be more significant.  We typically have $H \ll m_{\rm eff} \ll \Gamma_\nu, y_\f \bar \f$.
In this limit, which interests us, the ``measurement" of neutrino flavor occurs on a timescale much faster than the $\f$ oscillation timescale.
We can consider the system evolves adiabatically during the $\f$ oscillation. 
Then, the production of $N$ can be estimated by averaging the production rate over time, given by
\beq 
\G_{N}= \G_\nu \overline{\sin^2\(2\theta_{\rm eff}\)}.
\eeq 
The overline denotes the time average over the period for $\f$ oscillation during which we can neglect the Hubble expansion. 
 Here $\sin^2 \(2\theta_{\rm eff}\)$ is the oscillating effective mixing term
  \beq 
\sin^2 \(2\theta_{\rm eff}\)= \frac{(y_N v)^2 ((y_\f\f)^2+M_N^2) }{(y_N v)^2
((y_\f\f)^2+M_N^2)+((y_\f \f)^2+M_N^2
-
{2p}V_\nu)^2}.
 \eeq 
 Here we neglect the imaginary part contribution in e.g.~\cite{Abazajian:2001nj}. 
By using $\f =\bar{\f}\cos[m_{\rm eff}t]$, the time average gives 
\beq 
\overline{\sin^2(2\theta_{\rm eff})}\simeq \frac{\(y_N v\)^2}{2\sqrt{-2 p V_\nu+M_N^2} y_\f\bar{\f}}\(1+\frac{M_N^2}{-2 p V_\nu+M_N^2}\).
\eeq 
We have checked that this analytic formula agrees well with a numerical simulation solving a kinetic equation. The mixing effect becomes most important when $y_\f \bar{\phi}\ll M_N^2-2p V_\nu$ but this happens in a small fraction of time $\sim \sqrt{M_N^2-2p V_\nu}/(y_\f\bar \f)$ giving the suppression of the mixing. 
This highly suppresses the production of $N$ compared with the DW mechanism. 
One notices that $\G_N$ scales with $a^{-1}$ and dominates at the lower temperature in the radiation-dominated Universe until the bare mass effect of $M_N$ becomes important. {Therefore, the production is dominant at $-2p V_\nu \sim M_N^2$, which is similar to the condition in the original DW mechanism. Thus we adopt the cosmic temperature for the dominant DW production, $T\sim 145\MEV$ for keV sterile neutrino (see e.g. \cite{Gelmini:2019wfp}). 
{Thus we will focus on the production of sterile neutrino at this moment by noting that
at} this moment $\sin^{2}(2\theta_{\rm eff})\sim \sin^2(2\theta)$ for the original DW mechanism.

 In our case, at this moment, we still have a further suppression of $M_N/(y_\f \bar{\f})$ which can be as small as $\frac{M_N}{\KEV}\frac{100\MEV}{y_\f \bar\f}\sim 10^{-5}$. 
We can estimate the thermally produced $N,\bar N$ by using 
\beq 
\Omega^{\rm th}_N h^2 \sim \left.\frac{\G_\nu \overline{\sin^2(2\theta_{\rm eff})}}{H} \frac{T^3 M_N}{\pi^2s}\right|_{T=145\MEV} \frac{s_0}{\rho_c/h^2}=0.01\frac{145\MEV}{y_\f \bar{\f}} \(\frac{M_N}{1\KEV} \)^2 \(\frac{\theta}{0.05}\)^2,
\eeq 
where we used $y_N v\sim M_N\theta,$ and taken $p\sim T$. As a result, we can significantly suppress thermal production.
{We note that when the onset of oscillation happens at $T< 145\MEV,$  the effective mixing at $T=145\MEV$ is $\sim \frac{(y_N v)^2 }{(y_\f \f_{\rm ini})^2}$ and can be more suppressed.}

Production via the Shi-Fuller mechanism~\cite{Shi:1998km} is also highly suppressed because the production of the active neutrino asymmetry is also suppressed by  $\overline{\sin^2[2\theta_{\rm eff}]}$, which is also numerically checked.
Analytically, one can conservatively use $n_{N} \frac{\G_N}{H}|_{T=145\MEV}$ to estimate the asymmetric part of $\nu$ from the $N$ scattering via mixing. We find that the contribution to the thermal potential of $\nu$ is smaller than either the bare mass contribution or the symmetric contribution $\sim 10 G_F^2 T^5$ for the parameter of interest. Moreover, since $|\mu|\lesssim T$, we have Pauli blocking effect in this $N\to \nu$ process, and the asymmetry production, and thus the thermal potential contribution is further suppressed.  Therefore resonant conversion does not occur.\footnote{Given enough amount of active neutrino asymmetry by hand, we find a new phenomenon that the resonant conversion between active and sterile neutrino occurs multiply. This is because in this case we have a different sign in the thermal potential and during the oscillation of
$\f$, level-crossing occurs multiple times. }

Later, when $y_\f\bar{\f}$ redshifts and becomes smaller than $M_N$, the mass of $N$ becomes constant. This is a late-forming DM and  the DM needs to become ``matter" before $T\gtrsim 1\KEV$~\cite{Das:2020nwc}. This is satisfied if $y_\phi\bar\f\lesssim T$ at $T=145\MEV$.

\section{More RHNs and Seesaw
\label{sec:neutrinomass}}
By taking into account this suppression mechanism, we can extend the scenario with additional $n$ right-handed neutrinos. 
The right-handed neutrinos, $N_{i}$, with $i=1\cdots n$, can explain the neutrino oscillation scales $0.001-0.05$ eV via the seesaw mechanism.
As we have noted, the AWDM $N$ cannot explain the active neutrino mass scale because it has a Dirac mass rather than a Majorana mass. Suppose that initially not only $N$ but also $N_i$ carries the similar asymmetry, which may be from the decay of a single Affleck-Dine field. Then $N_i$ will also contribute to the baryon asymmetry. 
For simplicity, assuming that $N_i$ have Yukawa couplings of similar magnitude to the SM Higgs and leptons as $N$ does, it can enhance the relation~\Eq{asym} by a factor of $\sim n+1$.
We consider the following two scenarios. 

\paragraph{Scenario1: $M_{N_i}<M_N<\KEV$}
If $N_i$ are also DM, they do not contribute to the homochirality of amino acids because $N_i$ do not decay into circulary polarized photons due to their Majorana nature.

Then we may consider that $N_i$ are lighter than $N$, and since they have similar number densities, $N_i$ become subdominant DM. In this case, \Eq{DMasym} is enhanced by $\sim n+1$. We can avoid the thermal production of $N_i$ by coupling them to the same $\phi$ with a similar coupling strength as $N$.

\paragraph{Scenario2: $M_N<\KEV<M_{N_i}$ with $N_i$ decaying into dark radiation.}
On the other hand, we can also consider that $N_i$ decay much before the matter-radiation equality epoch, so they are not DM.
This can be possible if we also introduce new moduli to avoid the thermal production of $N_i$, while we do not require \Eq{yphi} for $N_i$.\footnote{We do not consider using the same modulus $\phi$ because, for $N_i$ to decay, the coupling should be large. Then the onset of $\phi$ oscillation may happen too early to have a large enough effective mass to suppress $N$ thermal production.}
Then
$N_i \to \nu \phi$ occurs much before the matter-radiation equality, which means \Eq{DMasym} is enhanced by $\sim n+1$. The energy density stored in $N_i$ becomes dark radiation, which is subdominant as long as the decay is not too late. This dark radiation may be probed by future cosmic microwave background experiments or baryonic acoustic oscillation experiments. The active neutrinos are produced from the decay and the certain cosmic neutrino background experiments may also probe this scenario.
\\

In both cases, we find a new coincidence. The allowed parameter region is shown in Fig.~\ref{fig:2} for $n=2$ explaining the neutrino oscillation scales. The two right-handed neutrinos are on the green bands.
In the region with $M_{N,N_i} < \text{keV}$, we assume that $N_i$ is lighter than $N$ and the dominant DM is $N$.
In the light gray region, we assume that they decay much before the matter-radiation equality into dark components, neutrino and $\phi$, so the constraints from X-ray do not apply.
In either case, we can see the parameter region for baryon asymmetry can agree with the seesaw relation when the two additional RHNs are also around the keV mass scale.

Interestingly, three right-handed neutrinos, including the Dirac one, with similar mass and similar Yukawa couplings, may simultaneously explain DM, baryon asymmetry, homochirality of amino acids, and neutrino mass.

\begin{figure}[t!]
\begin{center}  
\includegraphics[width=135mm]{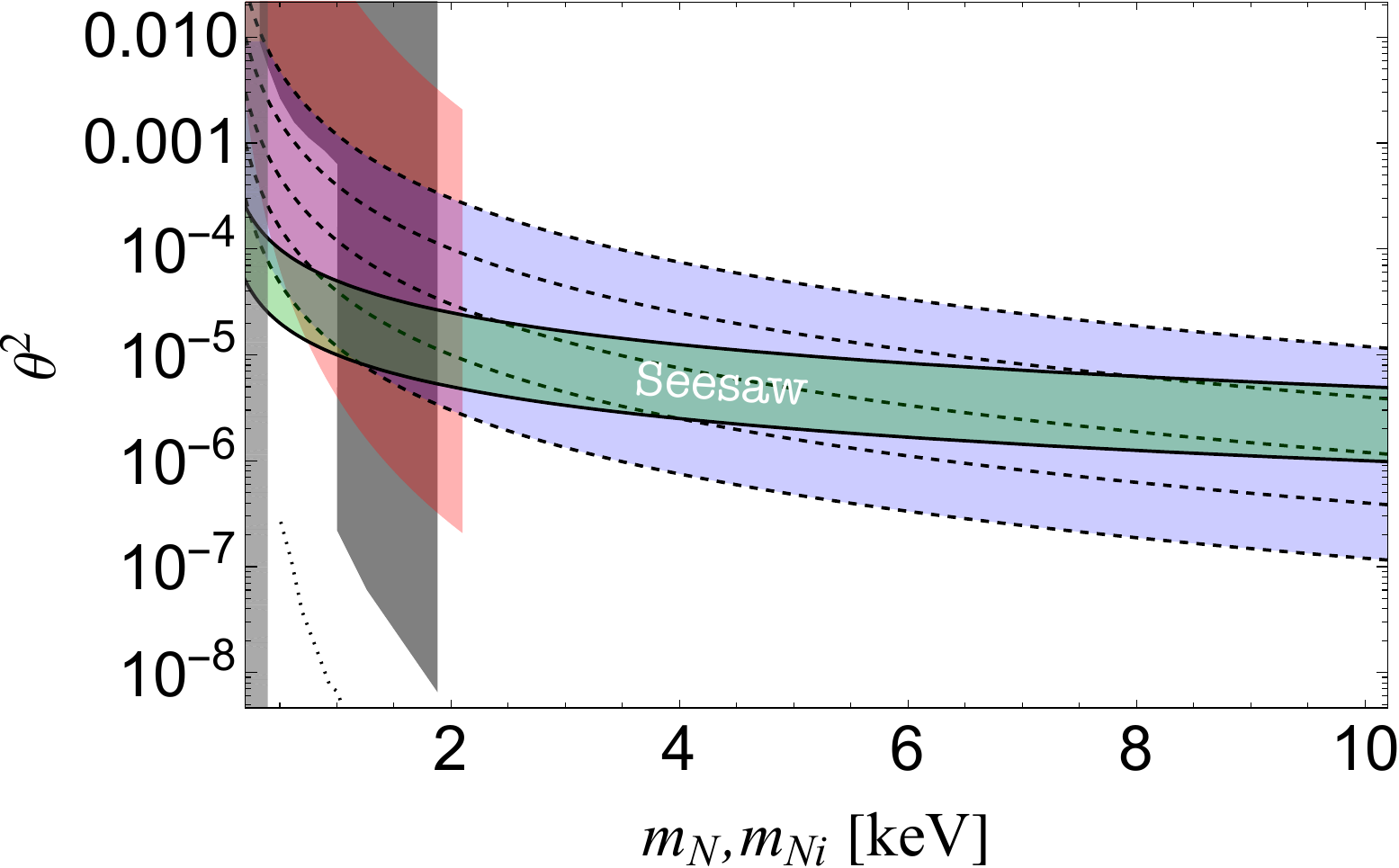}
\end{center}
\caption{The parameter region with three RHNs (including the AWDM) for explaining the neutrino mass scales. 
We assume that the two RHNs have the same size of Yukawa coupling as $N$ and have the same initial asymmetry.
This figure is valid for $M_{N_i}< M_N< \text{keV}$, or $M_{N_i}> \text{keV} > M_N$ with $N_i$ decaying into dark radiation well before matter-radiation equality.
 The three RHNs are on a single dashed line in the blue shaded region. 
The green band denotes the seesaw relation applying only to $N_i$. The gray, light gray, and red regions, as well as the dotted line, denote the same regions as in Fig.\,\ref{fig:1} and only apply to $N$, which is the dominant DM.   
Since X-ray constraints do not apply to $N_i$ decaying into dark radiation, the X-ray bound above  a few keV is removed.
}
\label{fig:2}
\end{figure}

\section{Conclusions and discussion
\label{sec:conclusions}}
We have explored a novel scenario involving asymmetric dark matter in the keV range, specifically focusing on right-handed or sterile neutrinos. We proposed that, according to the Fermi-Dirac distribution, this DM forms a Fermi degenerate gas and is consequently colder than dark matter produced through conventional thermal processes. Our model suggests that the tiny Yukawa interactions between the Standard Model lepton and the Higgs field could account for the baryon asymmetry of the Universe. Additionally, we hypothesized that the asymmetry between the left-handed and right-handed optical isomers of amino acids could result from the dark matter decay into circularly polarized photons. The viability of this scenario can be tested through ongoing observations of X-ray, eROSITA and XRISM, and future missions such as Athena and Theseus, with particular emphasis on measuring the polarization of the photons, like in IXPE \cite{10.1117/1.JATIS.8.2.026002}, to shed light on the origin of life. To further bolster our scenario, we propose a new mechanism to suppress the thermal production of DM by introducing a light modulus, which may also be beneficial for a successful cosmology involving generic right-handed neutrinos with large mixing.

This scenario also predicts that the asymmetry is global i.e. all over the Universe we are likely to have the same homochirality. Thus, another prediction is that future searches for the homochirality of amino acids in space will show a bias towards a single chirality.

\section*{Acknowledgement}
We thank Junichi Takahashi and Kensei Kobayashi for useful discussions on the homochirality of amino acids. 
This work was supported by JSPS KAKENHI Grant Nos.  20H05851(W.Y.), 21K20364(W.Y.), 22K14029(W.Y.), and 22H01215(W.Y.).
We thank OpenAI's ChatGPT-4 and DALL-E for generating the illustrations in Fig.~\ref{fig:0}. 
We also acknowledge the use of ChatGPT-4(o) and Bing for basic knowledge of life science and for searching the references related to it.
\appendix

\section{A model of large $N$ asymmetry} \lac{AD}
In our model, a large asymmetry of DM is a fundamental component for explaining cold DM, homochirality, and baryon asymmetry.
For a mechanism creating the asymmetry, we can consider the decay of a complex scalar field with lepton number, $q$, denoted by $\F$,  which is the Affleck-Dine field~\cite{Affleck:1984fy,Murayama:1993em,Dine:1995kz}. We do not discuss a concrete model or potential shape for the Affleck-Dine mechanism in detail since it has already been widely studied. Let us discuss the cosmology instead. 
While $\F$ can acquire a large expectation value in the presence of Hubble-induced operators during inflation, it starts to oscillate around the origin of the potential.
We consider a potential that generates the rotation of the phase of $\F$ in a CP-violating way. Then we get the lepton number density, $n_L=iq(\dot{\F}^*\F-\F^*\dot{\F})$, with a sign determined by $q$ and the direction of rotation.
The lepton asymmetric number density carried by the condensate can be $\mathcal{O}(1)$ if the rotation is dominant compared with the radial motion.
To generate the $\bar{N}$ asymmetry, we can couple $\F$ to $N$ in a lepton-number-conserving way, e.g.
\beq 
{\cal L}_{{\rm int},N}= -y_\F \F \bar{N}^c N,
\eeq 
where $y_\F$ is a Yukawa coupling.
If the mass of $\F$ is larger than the twice of the mass of $N$,  the decay can happen.

Then we can 
produce the lepton asymmetry of $N$ with
\beq
\e \equiv \frac{ n_N-n_{\bar{N}}}{n_N+n_{\bar N}}\sim -\O(1).
\eeq
In particular, this is the case that $\F$ decay ends with a cosmic-temperature similar to the mass of $\F$ (see e.g. \cite{Moroi:2020has,Moroi:2020bkq}).\footnote{
In the scenario for suppressing the thermal production
by the modulus,  we had  $y_\f \f_{\rm amp}\lesssim T_{\rm osc} $ at the onset of oscillation of $\f$ and assumed that the production of $N$ asymmetry, i.e. the decay of $\F$, is much earlier than the onset of oscillation. 
Thus at the end of $\F$ decay, $T>T_{\rm osc}$, the effective mass of $N$ is $\sim y_\f \f_{\rm ini} \ll T$. This is much smaller than the mass of $\F$ if it is around $T$. }

\section{Cases with $m_{\rm eff} \gg \Gamma_\nu$} \lac{opposite}
In Sec.\ref{sec:suppression}, we focused on the case of $m_{\rm eff}\ll \Gamma_\n$, and showed that the effective mixing coupling is significantly suppressed.
In the opposite limit, the mixing term is also suppressed.
Before the ``measurent" of the flavor, the $\f$ oscillation effect is important. 
At the amplitude level, we can take the time average of $M^\dagger \cdot M/(2p)$ over the mean free time $1/\Gamma_\nu$. The time average of the off-diagonal component of $M^\dagger \cdot M/(2p)$ 
\beq \sim \frac{\O(y_N v y \bar \f)}{p} \times \frac{\Gamma_\nu}{m_{\rm eff}}.\eeq
 Here we use the fact that the time integral is destructive for time scales longer than $1/m_{\rm eff}$.
The time average of the $(N,N)$ component is 
$\O((y \bar \f)^2/p)$ by neglecting $M_N$. This is constructive. 
Therefore, the mixing is effectively suppressed by the ratio:
\beq 
\sin{\theta_{\rm eff}}\sim \O\(\frac{y_N v}{y \bar \f} \) \frac{\Gamma_\nu}{m_{\rm eff}}.
\eeq 

One can see that when $m_{\rm eff} \gg \Gamma_\nu$, we can also significantly suppress the $N$ thermal production. This suppression is also observed in a numerical simulation. The mechanism can be useful in a generic scenario that requires the suppression of the thermal production of RHNs.

\bibliography{reference}

\end{document}